\documentclass[preprint,groupedaddress,floatfix,%
nofootinbib]{revtex4}

\usepackage{euscript}
\usepackage{dcolumn}
\usepackage{graphicx}
\usepackage{epsfig}
\usepackage{hyperref}
\usepackage{graphicx}
\usepackage{dcolumn}
\usepackage{longtable}
\usepackage{times}
\usepackage{amsmath,amssymb,bm}
\usepackage[english]{babel}
\usepackage[latin1]{inputenc}
\usepackage{times}
\usepackage[T1]{fontenc}
\usepackage{color}
\usepackage{fancybox}
\usepackage{sidecap}
\usepackage{graphicx}
\usepackage{epsfig}
\usepackage{psfrag}
\usepackage{bbm}
\usepackage[english]{babel}
\usepackage[latin1]{inputenc}
\usepackage{times}
\usepackage[T1]{fontenc}
\usepackage{color}
\usepackage{fancybox}
\usepackage{sidecap}
\usepackage{psfrag}
\usepackage{bbm}
\usepackage{wasysym}
\usepackage{slashed}
\usepackage{tikz}
\usepackage{euscript}

\newcommand{\mathi}{\mathrm{i}}
\newcommand{\mathe}{\mathrm{e}}

\begin{document}

\title{Testing APS conjecture on regular graphs}
\author{Wenxuan Tao}
\author{Fen Zuo\footnote{Email: \textsf{zuofen@miqroera.com}}}
\affiliation{Shanghai MiQro Era Digital Technology Co. Ltd., Shanghai, China}

\begin{abstract}
The maximum energy of the EPR model on a weighted graph is known to be upper-bounded by the sum of the total weight and the value of maximum-weight fractional matching~(MWFM). Recently, Apte, Parekh and Sud~(APS) conjecture that the bound could be strengthened by replacing MWFM with maximum weight matching~(MWM). Here we test this conjecture on a special class of regular graphs that Henning and Yeo constructed many years ago. On this class of regular graphs, MWMs achieve tight lower bounds. As for the maximum energy of the EPR model, we have recently devised a new algorithm called Fractional Entanglement Distribution~(FED) based on quasi-homogeneous fractional matchings, which could achieve rather high accuracy. Applying the FED algorithm to the EPR model on Henning-Yeo graphs, we could thus obtain energy as high as possible and matching value as low as possible, and then make high-precision tests of the APS conjecture. Nevertheless, our numerical results do not show any evidence that the APS conjecture could be violated.

\end{abstract}

\maketitle

\tableofcontents

\section{Introduction}

For the classical Maximum Cut problem, we have the famous Goemans-Williamson algorithm based on semidefinite programming, which could achieve an approximation ratio at least .878~\cite{GW-1995}. Quantum Maximum Cut~(QMC) is the natural quantum lift of the classical problem, in which adjacent qubits prefer antisymmetric singlet pairing. For a weighted graph~$G(V,E,w)$, its Hamiltonian is defined as
\begin{equation}
H^\textrm{QMC}_G= \sum_{(ij)} w_{ij}h_{ij}\equiv\frac{1}{2}\sum_{(ij)} w_{ij}(I_iI_j-X_iX_j-Z_iZ_j-Y_iY_j),
\end{equation}

We could also have symmetric pairings, and the corresponding model is called Einstein-Podolsky-Rosen~(EPR) model~\cite{King-2023}, with the Hamiltonian:
 \begin{equation}
H^\textrm{EPR}_G= \sum_{(ij)} w_{ij}g_{ij}\equiv\frac{1}{2}\sum_{(ij)} w_{ij}(I_iI_j+X_iX_j+Z_iZ_j-Y_iY_j).
\end{equation}

Naturally, people generalize the classical semi-definite programming and develop the quantum version of it, and obtain a whole hierarchy of semidefinite programmings, the so-called quantum Lasserre hierarchy, also called Navascues-Pironio-Ac\'{i}n~(NPA) hierarchy~\cite{Lasserre-2002, NPA-2008, BH-2016}. The quantum Lasserre hierarchy could be rather complicated, especially when the level is not small. Nevertheless, it could provide us with much useful information. For example, at level 2, quantum Lasserre hierarchy could be used to prove that the maximum energy of both QMC and EPR model has the following upper bound~\cite{LP-2024}:
\begin{equation}
\lambda_\textrm{max}(H_G)\le w(G)+w(\textrm{FM}_G),\label{eq.FMb}
\end{equation}
where $w(G)$ is the total weight of the graph, $w(\textrm{FM}_G)$ is the value of the Maximum-Weight Fractional Matching~(MWFM).
In some sense, such an upper bound from MWFM seems to reflect monogamy of entanglement~\cite{CKW-1999}.

Recently, Apte, Parekn and Sud utilize the relation between token graphs and quantum many-body models to conjecture that, this upper bound could be strengthened by replacing MWFM with the Maximum-Weight Matching~(MWM)~\cite{APS-2025}£º
\begin{equation}
\lambda_\textrm{max}(H_G)\le w(G)+w(\textrm{M}_G),\label{eq.Mb}
\end{equation}
where $w(\textrm{M}_G)$ is the value of MWM. As the maximum energy of the EPR model is shown to be always larger than that of QMC~\cite{APS-2025}, we focus our attention on EPR, and call this strengthened statement for EPR the APS conjecture.
Up to now, it seems not easy to prove APS conjecture directly with quantum Lasserre hierarchy. Some preliminary attempts have been made in~\cite{LP-2024}\cite{JKKSW-2024}\cite{GSS-2025}.

Now we know that, APS conjecture is valid on complete graphs and cycle graphs~\cite{APS-2025}. Moreover, for all unweighted graphs and some weighted graphs with order not larger than 10, the conjecture is also valid. Naturally we would like to test it on more complicated graphs, and regular graphs provide nice options. On the one hand, regular graphs possess natural perfect fractional matchings, while maximum matching could differ from perfect matching a lot~\cite{Henning-2007}. On the other hand, people have  recently devised good approximation algorithms for the EPR model~\cite{JN-2025}\cite{ALMPS-2025}. The latter, which we call the ALMPS algorithm, has been further refined by us into the so-called Fractional Entanglement Distribution~(FED) algorithm~\cite{TZ-2025}. FED could achieve rather high approximation ratios on regular graphs~\cite{TZ-2025}. Therefore, we could perform very accurate tests for the APS conjecture on specific regular graphs. This would be the aim of this paper.

The remaining part of the paper is organized as follows. In the next section, we introduce the ALMPS and FED algorithms for the EPR model. In section III we review the Henning-Yeo construction for specific regular graphs. Then in section IV FED is applied to Henning-Yeo regular graphs to test the APS conjecture. In the final section we summarize the results and make a discussion.

\section{Approximation algorithms for the EPR model}

We start with the EPR model, which was introduced in \cite{King-2023}. We rewrite the Hamiltonian here
 \begin{equation}
H^\textrm{EPR}_G= \sum_{(ij)} w_{ij}g_{ij}\equiv\frac{1}{2}\sum_{(ij)} w_{ij}(I_iI_j+X_iX_j+Z_iZ_j-Y_iY_j).
\end{equation}
In fact, after introducing the model King immediately conceived a framework to approximate its maximum energy. This framework is based on the following generalized graph states, which we call magic graph states:
\begin{equation}
|\chi\rangle \equiv\prod_{(ij)}\mathe^{\mathi \theta_{ij}P_iP_j} |0\rangle^{\otimes n},\label{eq.chi}
\end{equation}
where
\begin{equation}
P=T Y T^\dagger =\frac{1}{\sqrt{2}}(X-Y).
\end{equation}
Notice that, if we further define
\begin{equation}
Q=T X T^\dagger=\frac{1}{\sqrt{2}}(X+Y),
\end{equation}
then
\begin{equation}
X_iX_j-Y_iY_j=P_iQ_j+Q_iP_j.
\end{equation}
Now we can calculate the expectation values of the EPR Hamiltonian on the magic graph state, which have been done explicitly in \cite{King-2023}. We collect the final expressions here:
\begin{eqnarray}
\langle\chi| Q_iP_j|\chi\rangle &=&\sin 2\theta_{ij} \prod_{k\in K} \cos 2\theta_{ik},\label{eq.QP}\\
\langle\chi| P_iQ_j|\chi\rangle &=&\sin 2\theta_{ij} \prod_{l\in L} \cos 2\theta_{jl},\label{eq.PQ}\\
\langle \chi| Z_iZ_j|\chi\rangle &=&\sum_{S\subset T, |S|\textrm{even}} \prod_{s\in S}\sin 2\theta_{is}\sin 2\theta_{sj} \prod_{k\in K\setminus S} \cos 2\theta_{ik}\prod_{l\in L\setminus S} \cos 2\theta_{lj}.\label{eq.ZZ}
\end{eqnarray}
Here
\begin{equation}
K\equiv \{k\in V_G: k\sim i, k\ne j\}, \quad L\equiv \{l\in V_G: l\sim j, l\ne i\},
\end{equation}
and
\begin{equation}
T\equiv\{t\in V_G: t\sim i, t\sim j\}.
\end{equation}
Notice that the expression $\langle \chi| Z_iZ_j|\chi\rangle$ is rather complicated. However, if the rotation angles on those edges incident to $T$ are identical, it could be greatly simplified, see Appendix \ref{APP.A} and Appendix \ref{APP.B}. The two algorithms discussed below heavily rely on the above expressions for the expectation values.

\subsection{ALMPS algorithm}

Based on the upper bound from MWFM~(\ref{eq.FMb}), \cite{JN-2025} and \cite{ALMPS-2025} independently develop algorithms for the EPR model with identical approximation ratio $.809$. Here we focus on the ALMPS algorithm~\cite{ALMPS-2025}. The main idea is to choose the rotation angles in the magic graph state (\ref{eq.chi}) based on MWFM. If the MWFM is given by $\{m_{ij}|(ij)\in E(G)\}$, then the corresponding angles are chosen as:
\begin{equation}
\cos 2\theta_{ij}=\exp[-\kappa\cdot m_{ij}], \quad \theta_{ij}\in [0,\pi/4],
\end{equation}
where $\kappa$ is a positive constant. With such a choice, we could simplify (\ref{eq.QP}-\ref{eq.ZZ}) by using the property of fractional matching. We could convert all the other angles in (\ref{eq.QP}) and (\ref{eq.PQ}) back into the current angle $\theta_{ij}$, or $m_{ij}$ instead. As for (\ref{eq.ZZ}), all the higher terms involve sines, which would be small and neglected when the angles are small; the leading term could be dealt with in the same way as (\ref{eq.QP}) and (\ref{eq.PQ}), and expressed solely with $\theta_{ij}$ or $m_{ij}$. Therefore, one obtains the following expression for the energy~\cite{ALMPS-2025}:
\begin{equation}
\langle \chi| g_{ij}|\chi\rangle \ge T(\kappa,m_{ij})\equiv \frac{1}{2}(1+\exp[-2\kappa(1-m_{ij})]+2\sqrt{1-\exp[-2\kappa m_{ij}]}\exp[-\kappa(1-m_{ij})]).
\end{equation}

Considering the fractional matching bound (\ref{eq.FMb}), the performance of the algorithm on an individual edge would rely on the following ratio
\begin{equation}
R(\kappa,m_{ij})\equiv T(\kappa,m_{ij})/(1+m_{ij}).
\end{equation}
As $m_{ij}$ in a MWFM could take any value in $[0,1]$, we need to calculate the following Max-Min value:
\begin{equation}
r_{0}\equiv\max_{\kappa>0} \min _{x\in[0,1]} R(\kappa,x).\label{eq.r0}
\end{equation}
\cite{ALMPS-2025} proves that this is half the gold ratio $\varphi= (1+\sqrt{5})/2$:
\begin{equation}
r_{0}=\frac{\varphi}{2}\approx .809,
\end{equation}
and the parameter $\kappa$ is given by
\begin{equation}
\kappa_{0}\equiv \frac{1}{2}\ln \varphi \approx .240.
\end{equation}
With the energy ratio on individual edges, \cite{ALMPS-2025} then prove that the total energy could achieve the same approximation ratio
\begin{equation}
r(\textrm{ALMPS})\equiv \frac{\langle \chi|H^\textrm{EPR}_G|\chi\rangle_{\textrm{ALMPS}}}{w(G)+w(\textrm{FM}_G)}\ge r_{0}\approx .809.
\end{equation}
\cite{JN-2025} obtains the same approximation ratio with a different quantum state, and will not be discussed here.

\subsection{FED algorithm}
\label{subsection:FED}

Although the approximation ratio $.809$ obtained in \cite{JN-2025} and \cite{ALMPS-2025} is much larger than those in previous studies, it seems not good enough for testing the APS conjecture. In \cite{TZ-2025}, we have refined the ALMPS algorithm, by taking fractional matchings as homogeneous as possible. As a result, the quantum entanglement in the corresponding states would be distributed more uniformly. Thus we call the refined algorithm ``Fractional Entanglement Distribution'', FED for short.

Let us give more details. From the calculation of the Max-Min value (\ref{eq.r0}) we can see that, the approximation ratio of the ALMPS algorithm is severely restricted by the possible values of the matching fractions $m_{ij}$. If we allow it to take any values in $[0,1]$, then $r_0$ is the best that the ALMPS algorithm could achieve. One improving strategy could be, to shrink properly the fraction interval, so as to increase the value of (\ref{eq.r0}). Explicitly, if we set the interval of the matching fractions as $I_{\delta,\Delta}=[\Delta^{-1},\delta^{-1}]$, then (\ref{eq.r0}) becomes
\begin{equation}
r_{\delta,\Delta}=\max_{\kappa>0} \min _{x\in I_{\delta,\Delta}} R(\kappa,x). \label{eq.rdD}
\end{equation}
It is not difficult to see that, when $I_{\delta,\Delta}$ gradually shrinks, the above value would gradually increase. Of course, we can not arbitrarily set the interval. The choice should still be made to guarantee that we could have large enough fractional matchings. For a given interval $I_{\delta,\Delta}=[\Delta^{-1},\delta^{-1}]$, suppose we could manage to find the largest fractional matching $\textrm{FM}_G^{\delta,\Delta}$. We then define the following ratio
\begin{equation}
s^{\delta,\Delta}_G\equiv \frac{w(\textrm{FM}^{\delta,\Delta}_G)}{w(\textrm{FM}_G)},
\end{equation}
and use it as a criterion whether the interval has been chosen properly. In fact, for the EPR model it would be better if we consider the shifted ratio by the total weight:
\begin{equation}
\hat s^{\delta,\Delta}_G\equiv \frac{w(G)+w(\textrm{FM}^{\delta,\Delta}_G)}{w(G)+w(\textrm{FM}_G)}.
\end{equation}
With all these preparations, we could now use $r_{\delta,\Delta}$ and $\hat s^{\delta,\Delta}_G$ to characterize the refined approximation ratio~\cite{TZ-2025}:
\begin{equation}
r(\textrm{FED})\equiv \frac{\langle \chi|H^\textrm{EPR}_G|\chi\rangle_{\textrm{FED}}}{w(G)+w(\textrm{FM}_G)}\ge r_{\delta,\Delta}\cdot \hat s^{\delta,\Delta}_G.
\end{equation}
When the $I_{\delta,\Delta}$ is properly chosen and the fractional matching is carefully constructed, the above approximation ratio could be quite satisfying.

As a preliminary setup, in \cite{TZ-2025} we proposed to construct a proper fractional matching in the following way. First, we define the edge degree as
\begin{equation}
d_{ij}=\max\{d_i,d_j\}.
\end{equation}
Then, we set the matching fraction as
\begin{equation}
m_{ij}=1/d_{ij}, \quad \forall (ij)\in E_G.
\end{equation}
As a result, $\Delta$ would be the maximum edge degree, and $\delta$ the minimum one.
Furthermore, $\Delta$ would coincide with the maximum degree of vertices, and $\delta$ would be no less than the minimum degree of vertices. For weighted graphs, we first round all weights into integers properly, use multi-edges to represent the weights, and then make the above assignment. Of course, such a framework is very rough, and could be easily improved. Nevertheless, we show that such an approach could actually increase the approximation ratios for particular graphs. For example, for unweighted $k$-regular graphs, the matching interval would shrink into a single point $1/k$, while we still have perfect fractional matching. So we have $\hat s^{k,k}_G=1$. And as the matching fraction takes only a single value, (\ref{eq.rdD}) degenerates accordingly into
\begin{equation}
r_k\equiv r_{k,k}=\max_{\kappa>0} R(\kappa,1/k). \label{eq.rkk}
\end{equation}
For small values of $k$, $r_k$ could be easily calculated, and the results are shown in Tab.~\ref{Tab.Gd}.

\begin{table}[h]
\begin{center}
\begin{tabular}{|c||c||c|}
\hline
$\quad k \quad $  &  $\quad r_k \quad $ &   $\quad \kappa_k \quad $          \\
\hline\hline
2           &  .872  &  .324    \\
\hline
3        &  .894 &  .203    \\
\hline
4       &  .912 &  .147   \\
\hline
5         &  .924 &  .115   \\
\hline
6       &  .934 &  .0945    \\
\hline
7        &  .942 &  .080  \\
\hline
8         &  .948 &  .0692   \\
\hline
9     &  .953 &  .061 \\
\hline
10      &  .957 &  .0544\\
\hline
\end{tabular}
\\
\caption{Approximation ratios $r_k$ of the FED algorithm for the EPR model on $k$-regular graphs, together with the corresponding parameters $\kappa_k$.}
\label{Tab.Gd}
\end{center}
\end{table}

It is clear from Tab. \ref{Tab.Gd} that, the approximation ratios $r_k$ for $k$-regular graphs greatly improve over $r_0\approx .809$ of the ALMPS algorithm, and $r_k$ continues to increase as $k$ increases. As a result, we could expect to make high-precision tests of the APS conjecture on regular graphs.

\section{Henning-Yeo regular graphs}

Nevertheless, regular graphs are still too huge a class to start with. We could choose those specific ones which are most unfavourable for the APS conjecture, namely those achieving the tight lower bounds of maximum matchings. These tight lower bounds and the corresponding graphs achieving them are obtained concretely in \cite{Henning-2007}. We simply list these results from  \cite{Henning-2007} below.

\subsection{Tight Lower Bounds for Maximum Matchings}

For the maximum matchings $\textrm{M}_k$ of $k$-regular graphs $G_k$, \cite{Henning-2007} shows
\begin{eqnarray}
|\textrm{M}_k|&\ge&  \frac{k^2+4}{k^2+k+2}\cdot\frac{n}{2}, \quad  k\ge 4 \quad \mbox{even},\\
|\textrm{M}_k|&\ge& \frac{(k^3-k^2-2)n-2k+2}{2(k^3-3k)},  \quad k\ge 3 \quad \mbox{odd},
\end{eqnarray}
and all these lower bounds are tight. We denote the matchings achieving these lower bounds as $\underline{\textrm{M}}_k$. Since the maximum fractional matching is always perfect, we have $|\textrm{FM}_k|=n/2$. We then define the ratio between the values of the maximum matching and fractional matching as
\begin{equation}
m_k\equiv \frac{|\underline{\textrm{M}}_k|}{|\textrm{FM}_k|}.
\end{equation}
Then the above inequalities could be reexpressed as
\begin{eqnarray}
m_k&=&  \frac{k^2+4}{k^2+k+2},\quad  k\ge 4 \quad \mbox{even},\\
m_k&=& \frac{(k^3-k^2-2)}{(k^3-3k)}+O(1/n), \quad k\ge 3 \quad \mbox{odd}.
\end{eqnarray}
Notice that for both the even series and odd series of $k$, we have $m_k\to 1$ when $k\to \infty$. That is, when $k$ is large enough, all the maximum matchings would approach the perfect one. In the later analysis, what we really need is the sum of the matching value and the total edge number or size. As the size of an order $n$ $k$-regular graph is $\lfloor kn/2 \rfloor$, we define the shifted ratio of the maximum matchings as
\begin{equation}
\hat m_k \equiv \frac{\lfloor kn/2 \rfloor+ |\underline{\textrm{M}}_k|}{\lfloor kn/2 \rfloor+|\textrm{FM}_k|}.
\end{equation}
Taking $n$ large enough, we immediately obtain the numerical values of $m_k$ and $\hat m_k$. Their values for small $k$ are shown in Tab.~\ref{Tab.Mk}. From the table we can see, although $m_k$ differ slightly from $1$, the shifted ratios $\hat m_k$ are very close to $1$.

\begin{table}[h]
\begin{center}
\begin{tabular}{|c||c|c|}
\hline
$\quad k \quad $  & $\quad m_k  \quad $ &  $\quad \hat m_k \quad $          \\
\hline\hline
3       & .889  &.972    \\
\hline
4       & .909 &  .982   \\
\hline
5         &.891 &  .982  \\
\hline
6       & .909 & .987   \\
\hline
7        &.907 &  .988 \\
\hline
8        &.919 &  .991   \\
\hline
9        & .92 & .992 \\
\hline
10       & .9286 &  .9935\\
\hline
\end{tabular}
\\
\caption{The ratios $m_k$ between tight lower bounds of maximum matchings and perfect fractional matchings on $k$-regular graphs, together with the size-shifted ratios $\hat m_k$.}
\label{Tab.Mk}
\end{center}
\end{table}

\subsection{Henning-Yeo graphs}

In \cite{Henning-2007} the authors also give the concrete construction for the specific regular graphs achieving the tight lower bounds.
The underlying principle is the Tutte-Berge theorem~\cite{Tutte-1947,Berge-1958}. Explicitly, for even degree $k\ge 4$, one takes the following steps:\\
1. Take an arbitrary $k$-regular graph $X_p$, which could contain multi-edges;\\
2. Delete an arbitrary edge $xy$ from the complete graph $K_{k+1}$, and get quasi-complete graph $H_{k+1}^{x,y}$;\\
3. For each edge of $X_p$, insert a copy $H_{k+1}^{u',v'}$  of the quasi-complete graph $H_{k+1}^{x,y}$ in such a way: delete edge $uv$, add $uu'$ and $vv'$.

Label the resulting graph by $G_p^k$. Obviously its order is $|G_p^k|=p(k^2+k+2)/2$. Also it is not difficult to prove that its maximum matching achieves the lower bound. Just delete the vertex set of $X_p$ and apply the Tutte-Berge theorem.

For odd degree $k\ge 3$, the construction is slightly complicated, and goes like this:\\
1. Construct a bipartite graph $T_p^k$, such that one vertex set is $V_1=\{u_1,u_2,...,u_p\}$, another set is $V_2=\{v_1,v_2,...,v_{p(k-1)+1}\}$, and choose the neighbours of each $u_i$ as $N(u_i)=\{v_{(k-1)(i-1)+1},v_{(k-1)(i-1)+2}...,v_{(k-1)(i-1)+k}\}$;\\
2. Start with the complete graph $K_{k+2}$ with vertex set $\{w_1,w_2,...,w_{k+2}\}$. Delete the following edges to get a quasi-complete
graph $H_{k+1}^{w_{k+2}}$:
\begin{equation}
H^{w_{k+2}}_{k+2}=K_{k+2}-\left(\bigcup_{i=0}^{(k-3)/2}\{w_{2i+1}w_{2i+2}\}\right)\cup \{w_kw_{k+2},w_{k+1}w_{k+2}\}.
\end{equation}
3. For each vertex $v$ of $V_2$ with degree $d_v<k$, make $k-d_v$ copies $H_{k+2}^{x_j}$ of $H_{k+2}^{w_{k+2}}$, where $1\le j\le k-d_v$. Then connect $v$ with each $x_j$.

Denote the final graph as $H_p^k$. Its order is $|H_p^k|=n=p(k^3-3k)+k^2+2k+1$. One can check that its maximum matching also achieves the lower bound, by deleting $V_2$ and apply the Tutte-Berge theorem.

\section{Testing APS conjecture on Henning-Yeo graphs}

\subsection{Unweighted regular graphs}

Now we try to test the APS conjecture on the Henning-Yeo regular graphs. First, we gather the approximation ratios $r_k$ and the shifted matching ratios $\hat m_k$ together, and compare them in Tab.~\ref{Tab.HYk}.

\begin{table}[h]
\begin{center}
\begin{tabular}{|c||c||c|}
\hline
$\quad k \quad $  & $\quad r_k  \quad $ &  $\quad \hat m_k \quad $          \\
\hline\hline
3       & .894  &.972    \\
\hline
4       & .912 &  .982   \\
\hline
5         &.924 &  .982  \\
\hline
6       & .934 & .987   \\
\hline
7        &.942 &  .988 \\
\hline
8        &.948 &  .991   \\
\hline
9        & .953 & .992 \\
\hline
10       & .957 &  .993\\
\hline
\end{tabular}
\\
\caption{Comparison of the approximation ratios $r_k$ of FED on general regular graphs and the size-shifted matching ratios $\hat m_k$ on Henning-Yeo~$k$-regular graphs.}
\label{Tab.HYk}
\end{center}
\end{table}

From Tab.\ref{Tab.HYk} we see that, although the FED algorithm performs rather well on general regular graphs, its approximation ratios are still lower than the shifted matching ratios $\hat m_k$. However, the current ratios are for general regular graphs, and could be improved when applying to the specific Henning-Yeo graphs. Specifically, in the analyses of the original ALMPS algorithm and the refined FED algorithm, all the higher terms in the expression $\langle \chi| Z_iZ_j|\chi\rangle$ (\ref{eq.ZZ}) have been neglected. For particular graphs, we could manage to calculate the exact results, and thus improve the approximation. In particular, since the FED algorithm chooses the same angle on all the edges of regular graphs,  (\ref{eq.ZZ}) could be simplified greatly, as shown
in Appendix~\ref{APP.B}.

When the contributions of these higher terms are included, we improve the ratios $r_k$ into $\hat r_k$. For a fixed $k$, $\hat r_k$ varies very slightly for different instances of the Henning-Yeo graphs. Since the variations are very tiny, we simply ignore them. The final results for the improved ratios $\hat r_k$ are shown in Tab.~\ref{Tab.HYk2}, in comparison with the size-shifted matching ratios $\hat m_k$.

\begin{table}[h]
\begin{center}
\begin{tabular}{|c||c|c||c|}
\hline
$\quad k \quad $  & \quad $r_k$ \quad &$\quad \hat r_k  \quad $ &  $\quad \hat m_k \quad $          \\
\hline\hline
3       & .894  & .894  &.971    \\
\hline
4       & .912 &  .914 &  .982   \\
\hline
5         &.924 & .926  &   .982  \\
\hline
6       & .934 & .937 &  .987   \\
\hline
7        &.942 & .944 &  .988 \\
\hline
8        &.948 & .950 & .991   \\
\hline
9        & .953 & .954 &  .992 \\
\hline
10       & .957 & .9586 & .9935\\
\hline
\end{tabular}
\\
\caption{On unweighted Henning-Yeo $k$-regular graphs, the original approximation ratios $r_k$ of FED, the improved ones $\hat r_k$, and the size-shifted matching ratios $\hat m_k$ are compared.}
\label{Tab.HYk2}
\end{center}
\end{table}

Tab.~\ref{Tab.HYk2} clearly shows that, keeping those higher terms indeed increases the approximation ratios, but very slightly. For $3\le k\le 10$, the increases are about $.001-.002$, much smaller than the gap between $r_k$ and $\hat m_k$. Therefore, the improved ratios $\hat r_k$ are still lower than the matching ratios $\hat m_k$, giving no hints that the APS conjecture could be violated.

\subsection{Weighted regular graphs}

The shifted matching ratios $\hat m_k$ in Tab.~\ref{Tab.HYk2} are simply too high to approach.
This is because the size actually makes the dominant contribution, and the differences between matchings and fractional matchings become not significant any more. We could introduce different edge weights to change this situation. In the Henning-Yeo construction, lots of quasi-complete graphs $H_{k+1}^{x,y}$ and $H_{k+2}^{w_{k+2}}$ are used. When applying the Tutte-Berge theorem, they all become odd components, and require external edges to realize perfect matchings. We thus introduce larger weights in these odd components, and smaller weights on the external edges. As a result, the difference between MWMs and MWFMs would be enhanced, leading to smaller shifted ratios $\hat m_k$. Explicitly, we give all the internal edges of $H_{k+1}^{x,y}$ and $H_{k+2}^{w_{k+2}}$ one weight $w_1$, and the remaining edges another weight $w_2$. We then vary the weight ratio $d_w=w_1/w_2$ properly, and perform all the calculations again. Of course, on weighted graphs the matching ratio $m^w_k$ and shifted ratios $\hat m^w_k$ should be redefined accordingly:
\begin{eqnarray}
m^w_k &\equiv & \frac{w(\underline{\textrm{M}}^w_k)}{w(\textrm{FM}^w_k)},\\
\hat m^w_k & \equiv & \frac{w(G_k) + w(\underline{\textrm{M}}^w_k)}{w(G_k)+w(\textrm{FM}^w_k)}.
\end{eqnarray}

Furthermore, the original FED algorithm should be adjusted slightly. First we use the method proposed in \cite{TZ-2025} and describe in Subsection \ref{subsection:FED} to construct the fractional matching. For the present case, this gives rise to the following assignments of matching fractions and rotation angles:

1. When $k$ is even, we get one identical fraction for the internal edges of $H_{k+1}^{x,y}$, and another one the external edges. Therefore, we choose one rotation angle $\theta_1$ for the internal edges of $H_{k+1}^{x,y}$, and $\theta_2$ for the remaining ones;

2. When $k$ is odd, we get three different matching fractions: one for the internal edges of $H_{k+2}^x$, one for the external edges incident to $H_{k+2}^x$, and one for the other external ones. We assign $\theta_1$, $\theta_2$ and $\theta_3$ to them respectively.

 We then perform the remaining steps of the FED algorithm. Since the higher terms in $\langle \chi| Z_iZ_j|\chi\rangle$ (\ref{eq.ZZ}) only appear for the internal edges, we could still adopt the simplified formula in Appendix. \ref{APP.B} to do the calculation. We call the approximation ratios in the weighted case $\hat r^w_k$:
\begin{equation}
\hat r^w_k\equiv  \frac{\langle \chi|H^\textrm{EPR}_G|\chi\rangle_\textrm{FED}}{w(G_k)+w(\textrm{FM}^w_k)}.
\end{equation}

It turns out that, for a given $k$ $\hat r^w_k$ has very weak dependence on the choices of Henning-Yeo graphs. But when the weight ratio $d_w=w_1/w_2$ increases, $\hat r^w_k$ slightly decreases. Such a decrease almost vanishes when $d_w\ge 10$. Thus we choose $d_w=10$, calculate the values of $\hat r^w_k$, and compare them to the shifted matching ratios $\hat m^w_k$. The results are listed in Tab. \ref{Tab.HYk3}.

\begin{table}[h]
\begin{center}
\begin{tabular}{|c||c|c||c|c|}
\hline
$\quad k \quad $  & \quad $\hat r_k$ \quad &$\quad \hat r^w_k  \quad $ &  $\quad \hat m_k \quad $ &  $\quad \hat m^w_k \quad $          \\
\hline\hline
3       & .894  & .888 & .971  & .952  \\
\hline
4       & .914 &  .909 &  .982 & .962  \\
\hline
5         &.926&  .925  &   .982  & .977 \\
\hline
6       & .937 & .936 & .987  & .980  \\
\hline
7        &.944 & .943 &  .988 &  .986 \\
\hline
8        &.950 & .949 &  .991  &.988 \\
\hline
9        & .954 & .954 &  .992 & .991\\
\hline
10       & .9586 & .958 & .9935  &.992  \\
\hline
\end{tabular}
\\
\caption{On unweighted and weighted Henning-Yeo $k$-regular graphs, the improved approximation ratios $\hat r_k$ and $\hat r^w_k$ of the FED algorithm are compared with the shifted matching ratios $\hat m_k$ and $\hat m^w_k$ respectively.}
\label{Tab.HYk3}
\end{center}
\end{table}

Tab. \ref{Tab.HYk3} shows that, when proper weights are introduced, the shifted matching ratios indeed decrease as expected, by about $.001-.02$. But meanwhile, the approximation ratios also decrease slightly, by about $.001-.006$. When both effects are combined together, the gaps between $\hat r^w_k$ and $\hat m^w_k$ become smaller now. Unfortunately, this improvement is not enough for $\hat r^w_k$ to exceed $\hat m^w_k$, and we could draw no conclusion about the APS conjecture from these results. Nevertheless, the fact that the approximation ratios could remain nearly unchanged when the weights are included still reflect the nice performance of the FED algorithm anyhow.


\section{Discussion}

In this paper we attempt to test the APS conjecture for the EPR model with regular graphs. We gather two favorable conditions for this: first, we select the Henning-Yeo graphs to force the maximum matchings to take tight lower bounds, and thus enlarge the gaps between maximum matchings and fractional matchings; secondly, we adopt the recently proposed FED algorithm, which achieves very nice approximation ratios by distributing entanglement as uniformly as possible. Nevertheless, we do NOT find any hint that the APS conjecture could be violated. We then introduce larger weights for the odd components of Henning-Yeo graphs, in order to further enlarge the gaps between MWMs and MWFMs. This reduces the gaps between the approximate energies of the EPR model and the APS bound to some extent, but the conjecture still holds.

Yet we could analyze these results from a different point of view. They suggest that in order to distinguish the two bounds, approximation ratios of the adopted algorithms need to be very high. Tab. \ref{Tab.HYk3} shows that the approximation ratios need to be at least $.97$ for unweigthed regular graphs, and at least $.95$ for weighted ones. One can hardly believe that algorithms with such nice performance could indeed be found. For example, the analysis in~\cite{Hwang+-2021} demonstrates that assuming Unique Games Conjecture~\cite{Khot-2002}  and a further conjecture in Gaussian geometry, it is NP-hard to achieve a .956 approximation for QMC. Therefore, instead of devising better algorithms to test the conjecture, we would better accept its validity and study its consequences. For example, we could employ the conjecture to help us better understand the entanglement property of the quantum states. Furthermore, since the APS conjecture is a corollary of the token graph conjectures \cite{APS-2025}, it would also be interesting to investigate the consequences of these conjectures.


\newpage

\appendix

\section{King Formulae}
\label{APP.A}

First we repeat some definitions. On a weighted graph $G=\{V,E,w\}$, the Hamiltonian of the EPR model is defined as~\cite{King-2023A}:
 \begin{equation}
H^\textrm{EPR}_G= \sum_{(ij)} w_{ij}g_{ij}\equiv\frac{1}{2}\sum_{(ij)} w_{ij}(I_iI_j+X_iX_j+Z_iZ_j-Y_iY_j),
\end{equation}
where all the weights are positive.

To approximate its maximum-energy state, \cite{King-2023A} proposes to use the following magic graph states:
\begin{equation}
|\chi\rangle \equiv\prod_{(ij)}\mathe^{\mathi \theta_{ij}P_iP_j} |0\rangle^{\otimes n},
\end{equation}
where
\begin{equation}
P=T Y T^\dagger =\frac{1}{\sqrt{2}}(X-Y).
\end{equation}
Further defining
\begin{equation}
\quad Q=T X T^\dagger=\frac{1}{\sqrt{2}}(X+Y),
\end{equation}
we obtain
\begin{equation}
X_iX_j-Y_iY_j=P_iQ_j+Q_iP_j.
\end{equation}
Then we only need to calculate three expectation values, $\langle \chi|Q_iP_j|\chi\rangle$, $\langle \chi|P_iQ_j|\chi\rangle$, and $\langle \chi|Z_iZ_j|\chi\rangle$. \cite{King-2023A} gives the complete expressions in the general case:
\begin{eqnarray}
\langle\chi| Q_iP_j|\chi\rangle &=&\sin 2\theta_{ij} \prod_{k\in K} \cos 2\theta_{ik},\label{Aeq.QP}\\
\langle\chi| P_iQ_j|\chi\rangle &=&\sin 2\theta_{ij} \prod_{l\in L} \cos 2\theta_{jl},\label{Aeq.PQ}\\
\langle \chi| Z_iZ_j|\chi\rangle &=&\sum_{S\subset T, |S|\textrm{even}} \prod_{s\in S}\sin 2\theta_{is}\sin 2\theta_{sj} \prod_{k\in K\setminus S} \cos 2\theta_{ik}\prod_{l\in L\setminus S} \cos 2\theta_{lj}.\label{Aeq.ZZ}
\end{eqnarray}
where
\begin{equation}
K\equiv \{k\in V_G: k\sim i, k\ne j\}, \quad L\equiv \{l\in V_G: l\sim j, l\ne i\},
\end{equation}
and
\begin{equation}
T\equiv\{t\in V_G: t\sim i, t\sim j\}.
\end{equation}
It is clear that, the calculation of $\langle \chi|Z_iZ_j|\chi\rangle$ would be complicated, since we have to choose all the even subsets. When all the rotation angles are in the interval $[0,\pi/4]$, all terms in $\langle \chi|Z_iZ_j|\chi\rangle$ are positive. Furthermore, the powers of sines would be high if the cardinality of the subset $S$ is large, and thus the contributions would be rather small when the angles are small. Therefore, when analysing the algorithms we tend to neglect all the higher terms, and keep only the leading pure cosine term~\cite{ALMPS-2025A}\cite{TZ-2025A}. In principle this would weaken the performance. When we need very high accuracy, we hope to keep these higher terms. For general graphs, this would greatly increase the computing time. Nevertheless, if the graphs are very special, and our choices for the angles are also special, then we could greatly simplify the calculation of $\langle \chi|Z_iZ_j|\chi\rangle$, as shown in the following section.

\section{Simplified King Formula}
\label{APP.B}

\begin{figure}[h]
\centering
	\includegraphics[width=0.3\textwidth]{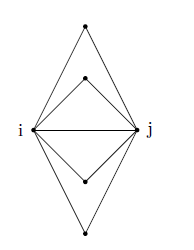}
\caption{For edge $ij$, the set $T$ composes of all common neighbours of $i$ and $j$.}\label{fig.triangle}
\end{figure}

The above expression tells us that, the complexity of $\langle \chi|Z_iZ_j|\chi\rangle$ originates from non-empty $T$ set, that is the set of shared neighbours of two adjacent vertices, as shown in Fig. \ref{fig.triangle}.

In one situation $\langle \chi|Z_iZ_j|\chi\rangle$ could be greatly simplified, that is when all the edges incident to $T$ are assigned identical rotation angles. Notice that we do not require the angles on the other edges of $i$ and $j$ to be the same. Therefore, we define
\begin{equation}
\tilde K\equiv K\setminus T,\quad \tilde L\equiv L\setminus T,
\end{equation}
and set
\begin{equation}
\tilde k=|\tilde K|,\quad \tilde l =|\tilde L|.
\end{equation}
Then we have
\begin{equation}
\langle \chi| Z_iZ_j|\chi\rangle = \prod_{k\in \tilde K} \cos 2\theta_{ik}\prod_{l\in \tilde L} \cos 2\theta_{lj}\sum_{S\subset T, |S|\textrm{even}} \prod_{s\in S}\sin 2\theta_{is}\sin 2\theta_{sj} \prod_{k\in T\setminus S} \cos 2\theta_{ik}\prod_{l\in T\setminus S} \cos 2\theta_{lj}.
\end{equation}
Notice that all the angles within the sum are identical, we could further simply it as
\begin{equation}
\langle \chi| Z_iZ_j|\chi\rangle = \prod_{k\in \tilde K} \cos 2\theta_{ik}\prod_{l\in \tilde L} \cos 2\theta_{lj}\sum_{S\subset T, |S|\textrm{even}} \sin ^ {2|S|} 2\theta \cos^{2|T|-2|S|} 2\theta.
\end{equation}
Now the summation only depends on the cardinalities of $T$ and $S$, and not on the details of the sets anymore. Denote $t=|T|$, $s=|S|$, we have
\begin{equation}
\langle \chi| Z_iZ_j|\chi\rangle = \prod_{k\in \tilde K} \cos 2\theta_{ik}\prod_{l\in \tilde L} \cos 2\theta_{lj}\sum^t_{s=0, s~\textrm{even}} \tbinom{t}{s}\sin ^ {2s} 2\theta \cos^{2t-2s} 2\theta.
\end{equation}
Using the binomial expansion, the summation could be calculated and leads to
\begin{equation}
\langle \chi| Z_iZ_j|\chi\rangle = \frac{1}{2}\left[1+(\cos^2 2\theta-\sin^2 2\theta)^t\right]\prod_{k\in \tilde K} \cos 2\theta_{ik}\prod_{l\in \tilde L} \cos 2\theta_{lj}.\label{Aeq.ZZ-simple}
\end{equation}
This is the final expression we want. It tells us that, for each edge $ij$, once we separate the individual neighbour sets $\tilde K$ and $\tilde L$ of $i$ and $j$ respectively, the remaining part only depends the cardinality $t$ of $T$ and could be calculated very efficiently. Now we use this simplified formula to compute the results for the complete graphs $K_n$ and quasi-complete graphs $H_{k+1}^{x,y}$ and $H_{k+2}^x$.

\subsection{Complete Graphs}

On complete graphs, every other vertex is adjacent to both $i$ and $j$, so $t=n-2$, and $\tilde K=\tilde L=\emptyset$. Thus
\begin{equation}
\langle \chi| Z_iZ_j|\chi\rangle =  \frac{1}{2}\left[1+(\cos^2 2\theta-\sin^2 2\theta)^{n-2}\right].
\end{equation}

\subsection{Quasi-Complete Graphs}

In the Henning-Yeo construction~\cite{Henning-2007A}, lots of quasi-complete graphs $H_{k+1}^{x,y}$ and $H_{k+2}^x$ are used. $H_{k+1}^{x,y}$ for even $k$ are very close to $K_{k+1}$, while $H_{k+2}^x$ for odd $k$ are very close to $K_{k+2}$. Now we calculate (\ref{Aeq.ZZ-simple}) for these two kinds of graphs.

Let us first review the construction of $H_{k+1}^{x,y}$. This is quite simple, we just remove the edge $xy$ from $K_{k+1}$. But considering $x$ and $y$ would be connected to external vertices $x'$ and $y'$ respectively, we include these two external edges into $H_{k+1}^{x,y}$. Then all $k+1$ vertices of $H_{k+1}^{x,y}$ would be of degree $k$. The calculation for the two external edges would be simple, so we only consider the internal edges. Of course, the effects from the two external edges should be kept. As for the angles, we has shown in the main text that we could assign an identical value $\theta$ for the angles on the internal edges identical value, while another value $\theta'$ for those on the external edges. Furthermore, the internal edges of $H_{k+1}^{x,y}$ could actually be separated into two classes: one of the form $xu$, where one endpoint is $x$ or $y$; the other edges would not be incident to $x$ or $y$. We deal with them separately:

For the edges of the form $xu$:
\begin{equation}
\tilde K=\{x'\} ~\textrm{or}~ \{y'\},\tilde L=\{y\} ~\textrm{or}~ \{x\}, \quad t=k-2.
\end{equation}
Thus,
\begin{equation}
\langle \chi| Z_xZ_u|\chi\rangle =  \frac{1}{2}\left[1+(\cos^2 2\theta-\sin^2 2\theta)^{k-2}\right]\cos 2\theta' \cos 2\theta.
\end{equation}
And for the remaining edges $uv$:
\begin{equation}
\tilde K=\tilde L=\emptyset, \quad t=k-1.
\end{equation}
So
\begin{equation}
\langle \chi| Z_uZ_v|\chi\rangle =  \frac{1}{2}\left[1+(\cos^2 2\theta-\sin^2 2\theta)^{k-1}\right].
\end{equation}

The graph $H_{k+2}^x$ could be dealt with similarly. Again denote the external vertex as $x'$, external angle as $\theta'$, and internal angle as $\theta$. The internal edges also separate into two classes, those incident to $x$, and the others.

For the edges $ux$ incident to $x$, we have
\begin{equation}
\tilde K=\{x', u-1 ~\textrm {or}~ u+1\} ,\tilde L=\{w_k,w_{k+1}\} , \quad t=k-3.
\end{equation}
So
\begin{equation}
\langle \chi| Z_xZ_u|\chi\rangle =  \frac{1}{2}\left[1+(\cos^2 2\theta-\sin^2 2\theta)^{k-3}\right]\cos 2\theta' \cos^3 2\theta.
\end{equation}
As for the remaining edges $uv$, the sets $\tilde K$ and $\tilde L$ will be slightly complicated, but their cardinalities are definite. This is enough. We have
\begin{equation}
\tilde k=\tilde l=1, \quad t=k-2.
\end{equation}
Thus
\begin{equation}
\langle \chi| Z_uZ_v|\chi\rangle = \frac{1}{2}\left[1+(\cos^2 2\theta-\sin^2 2\theta)^{k-2}\right]\cos^2 2\theta.
\end{equation}


\end{document}